\documentclass{iaus}
\usepackage{graphicx}

\title[Magnetic field and evolution] 
{The basic role of magnetic field in stellar evolution}

\author[Maeder et al.]   
{Andr\'e Maeder, Georges Meynet, Cyril Georgy, Sylvia Ekstr\"{o}m
}

\affiliation{Geneva Observatory, CH-1290 Sauverny,
Switzerland}

\pubyear{2009}
\volume{259}  
\pagerange{100-100}
\date{Dec. 10 2008 and in revised form ??}
\jname{Cosmic Magnetic Fields: from Planets to Stars 
and Galaxies}
\editors{K.G. Strassmeier, A.G. Kosovichev, J. Beckmann}
\begin{document}

\maketitle

\begin{abstract}
Magnetic field is playing an important role at all stages of star evolution from star formation to the endpoints. The main effects are briefly reviewed.
We also show that O--type stars have large convective envelopes, where convective dynamo could work. There, 
 fields in magnetostatic balance  have intensities of the order of 100 G.

A few OB stars with strong polar  fields (\cite[Henrichs et al. 2003a]{Henrichs03a})  show large
N--enhancements indicating  a strong internal mixing. We suggest that the meridional circulation enhanced by an internal rotation law close to uniform in
these magnetic stars  is responsible
for the observed mixing. Thus, it is not the magnetic field itself which makes the mixing, but the strong thermal instability associated to solid body rotation.

A critical question for evolution is whether a dynamo is at work in
radiative zones of rotating stars. The Tayler-Spruit (TS) dynamo is the
best candidate.  We derive some basic relations for dynamos
in radiative layers.  Evolutionary models with TS dynamo show important
effects: internal rotation coupling and enhanced mixing, all model outputs
being affected.
\end{abstract}

\firstsection 
\section{Introduction}

The magnetic field of a star is, like the scent of a flower, subtle and invisible,  but it plays an essential role in evolution.
Magnetic field  is often not accounted for in star models. However, the examples below  
show that from star formation to the endpoints as compact objects,  magnetic
fields and rotation  strongly influence the course of evolution and all  
model outputs.

In Sect. \ref{sectHR}, we give an overview where in evolution
 the fields are intervening. In Sect. \ref{sectHen}, we emphasize some 
critical observations. In Sect. \ref{sectdyn}, we focus on the general dynamo equations, with  examples in Sect. 5 and 6 for the Tayler--Spruit (TS) dynamo. 

\section{Overview on  the magnetic field in star formation and evolution} \label{sectHR}

\begin{figure}
\includegraphics[height=10cm,width=13cm,angle=00]{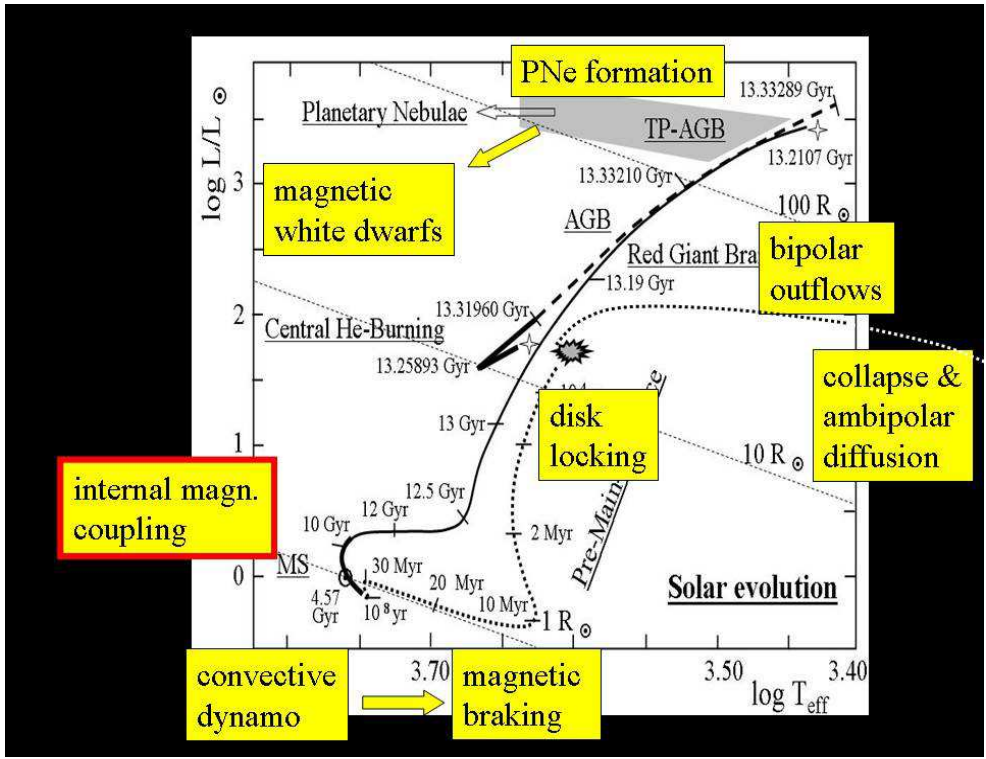}
  \caption{The evolutionary track of the Sun from the protostellar phase to the  phase of planetary nebulae  (courtesy from C. Charbonnel) with superposed indications of the various magnetic intervening processes}\label{Mfig1}
\end{figure}

Fig. \ref{Mfig1} shows the evolutionary track of the Sun from its formation
to its endpoint with indications of the various effects of the magnetic field coming into play.

 {\bf{--1. Collapse and ambipolar diffusion:}}
magnetic field may contribute to cloud support. For contraction to occur it is necessary that the energy density of magnetic field 
 $u_{\mathrm{B}}= {B^2}/({8 \; \pi})$ is smaller than the density of gravitational
 energy $u_{\mathrm{G}}= \frac{3}{5} \, \frac{GM^2}{R \;(\frac{4}{3} \pi R^3)} \;
= \; \frac{9}{20 \, \pi} \; \frac{GM^2}{R^4}$. This defines a critical mass $M_{\mathrm{B}}$ above which gravitation dominates,
\begin{eqnarray}
M_{\mathrm{B}} \approx \left( \frac{5}{18 \, \pi^2} \right)^{\frac{1}{2}} \frac{\Phi}{\sqrt{G}} \;
= \; 0.17 \, \frac{\Phi}{\sqrt{G}} \; ,
\end{eqnarray}
\noindent
with the magnetic flux $\Phi= \pi \, B \, R^2 $. In the original derivation 
(\cite[Mouschovias \& Spitzer 1976]{Mouschov76}), a numerical factor  0.13 was obtained instead of 0.17 in this simple derivation.
 If $M \, > \, M_{\mathrm{B}}$,
 large clusters or associations form. If $M \, < \, M_{\mathrm{B}}$, no contraction occurs until
the small ionized fraction ($\sim 10^{-7}$) to which the field is attached has 
diffused  (in about  $10^7$ yr)  out from the essentially neutral gas forming the cloud.

{\bf{--2. Bipolar outflows:}}
massive molecular outflows are  often detected in region of star formation.
A large fraction of the infalling material is not accreted by the central object
but it is ejected in the polar directions. In massive stars, the ejection cones are relatively broad.  Radiative heating and magnetic field are  likely driving the outflows. Remarkably the mass
outflow rates correlate with the luminosities of the central objects over 6 decades in luminosity,  from about
1 L$_{\odot}$ to $10^6$ L$_{\odot}$, as shown by \cite[Churchwell (1998)]{Church98}
and \cite[Henning et al. (2000)] {Henning00}.

{\bf{--3. Disk locking:}}
from the dense molecular clouds to the present Sun, the specific angular momentum
decreases by  $\sim 10^6$. Among the processes reducing the angular 
momentum in stars, disk locking is a major one (\cite[Hartmann 1998]{Hart98}). Fields of $\sim 1$ kG are sufficient for the coupling between the  star and a large accretion disk. The contracting star is bound to its disk and it keeps the same angular velocity during contraction, thus losing a lot of angular momentum.
The typical disk lifetime is a few $10^6$ yr.

{\bf{--4. Convective dynamos and magnetic braking:}}
solar types stars have external convective zones which produce a dynamo.
The resulting magnetic field creates a strong coupling between the star and the 
solar wind, which leads to losses of angular momentum. The relation expressing
these losses as a function of the stellar parameters have been developed by
\cite[Kawaler (1988)]{Kaw88}. Further improvements to account for saturation effects and 
mass dependence have been brought (\cite[Krishnamurti et al. 1997]{Krishna97}).

Massive OB stars also have significant external convective zones which may
represent up to 15\% of the radius. Surprisingly rotation enhances these convective zones (\cite[Maeder et al. 2008]{Maeder08}), which may also produce magnetic braking.

{\bf{--5. Dynamo in radiative zones:}}
is there a dynamo working in internal radiative zones? This is the biggest question concerning magnetic field and stellar evolution, with far reaching consequences
concerning mixing of the chemical elements and losses of the angular momentum.
This question is also essential regarding the rotation periods of pulsars and the origin
of GRBs. We devote Sect. 4 to 6 to this question.

{\bf{--6. Magnetic field in AGB stars, planetary nebulae and final stages:}}
red giants, AGB stars and supergiants have convective envelopes and thus dynamos.
Evidences of magnetic  fields up to kG in some central stars of planetary nebulae are given
(see Jordan, this meeting), they contribute to  shaping the nebulae (see Blackman, this meeting). White dwarfs have magnetic fields from about $10^4$  up to $10^9$ G,
the highest fields likely resulting from common envelope effects in cataclysmic variables.

\section{Magnetic fields and abundances: the Henrichs et al. results}  \label{sectHen}

 Since OB stars also have convective envelopes, the question arises what are the possible fields created by 
the associated dynamos. If one considers a flux tube in magnetostatic balance 
in the stellar  atmosphere, the equilibrium field is given by the condition
${B^2}/({8 \, \pi})  \, = \, P_{\mathrm{ext}}-P_{\mathrm{int}}$. At optical depth
$\tau=2/3$, the pressure is $P(\tau=2/3) \, \approx \, ({2}/{3})\, {g}/{\kappa}$.
Since $P_{\mathrm{int}}>0$, the maximum possible field $B_{\mathrm{eq}}$ for magnetic equilibrium is (\cite[Safier 1999]{Safier99})
 
\begin{eqnarray}
B_{\mathrm{eq}}\left(\tau={2}/{3}\right)\, \approx \, \left( \frac{16 \, \pi}{3} \, \frac{g}{\kappa} \right)^{1/2} \; .
\label{beq}
\end{eqnarray}
\noindent
\begin{table}[!th]
\caption{The equilibrium fields. The stars are on the ZAMS, except the Sun.} \label{tblfield}
\begin{center}\scriptsize
\begin{tabular}{rlrc}
            &             &                &       \\
 Spectral type  &  field  & Spectral type  & field \\              
\hline
            &             &           &   \\
      M0    & 2.8 kG      & G0        & 1.0 kG   \\
      KO    & 1.5 KG      & F2        & 0.6 kG    \\
 Sun        & 1.3 kG      & *O9        & 0.2 kG     \\
            &             &           &           \\
\hline
* from the author
\end{tabular}
\end{center}
\end{table}

\noindent

Table \ref{tblfield} gives the corresponding estimates for stars of various types.
 The observed field intensities  are often within a factor   of 2 from the 
 maximum  values given in the table. 
 Searches for magnetic fields in OB--type stars show no general evidence of fields above the  level of
$\sim$100 G  (\cite[Mathys 2004]{Mathys04}).  This is of the order of magnitude of the possible fields in magnetostatic equilibrium in the convective 
envelopes of OB stars. The strong fields of 1 kG or more
are not widespread (\cite[Hubrig et al. 2008] {Hubrig08}).
 
Noticeable exceptions were found and studied by Henrichs and colleagues
(\cite[Henrichs et al. 2003a]{Henrichs03a}, \cite[Henrichs et al. 2003b]{Henrichs03b}).  Their remarkable finding is that the few  stars with high
polar fields $B_{\mathrm{p}}$ also show N and He enhancements  together with C and O depletions, in particular for the 4 stars listed below.
  \begin{table}[!th]
  \caption{Stars with intense fields and N enrichments from (\cite[Henrichs, Neiner \& Geers 2003a]{Henrichs03a}) }
\begin{center}\scriptsize
\begin{tabular}{lllrlll}  \label{tblH}
               &         &            &     &            &          & \\
$\beta$    Cep & B1IVe   & v $\sin i=$& 27  &km s$^{-1}$ & $B_{\mathrm{p}}$=360 G & $\Delta \log$ N=1.2 \\
V2052    Oph & B1IVe   & v $\sin i=$& 63  &km s$^{-1}$ & $B_{\mathrm{p}}$=250 G &$\Delta \log$ N=1.3 \\
$\zeta$    Cas & B2IV    & v $\sin i=$& 17  &km s$^{-1}$ & $B_{\mathrm{p}}$=340 G&$\Delta \log$ N=2.6 \\
$\omega$ Ori & B2IVe   & v $\sin i=$& 172 &km s$^{-1}$ & $B_{\mathrm{p}}$=530 G &$\Delta \log$ N=1.8 \\
\end{tabular}
\end{center}
\end{table}
\noindent
The abundances are given in Table \ref{tblH} as the difference in log between the observed N abundance and the solar values. These  are typical signatures of CNO processing, which give strong evidences of internal
mixing in stars with a high magnetic field.  
These few results  bring a lot of interesting questions. 

The law of isorotation of Ferraro  clearly implies that an internal polar field enforces solid body rotation, see also 
Sect. \ref{num}. Now, the above results suggest that even in presence of
uniform rotation, there is an  efficient mixing. 
What is the  mixing process in stars with a polar magnetic field?
Shear turbulence is generally the main mixing process of chemical elements.
However, it is absent here, since the stars rotate uniformly.

The only process among those usually acting  in massive stars is meridional
circulation. Is it sufficient to produce such a mixing? In differentially 
rotating stars, it is usually much less efficient than shear mixing for the transport of the chemical
elements  (\cite[Meynet \& Maeder 2000]{MMV}), while it is very efficient for the transport of angular momentum.
However, we found that
meridional circulation is strongly enhanced by solid body rotation, since uniform
rotation creates a strong breakdown of radiative equilibrium.

 In evolutionary models with magnetic field and meridional circulation, there is a strong interplay between meridional circulation and magnetic field (\cite[Maeder \& Meynet 2005]{Magn3}):\\

	\begin{itemize}
 \item	Differential rotation creates the magnetic  field.
 \item  Magnetic field tends to suppress  differential rotation.
 \item  A rotation close to uniform strongly  enhances meridional circulation.
 \item  Meridional circulation  increases differential rotation and produces mixing.
 \item  Differential rotation feeds the dynamo and  magnetic field (the loop is closed).
 \end{itemize} 
 
 \vspace*{3mm}
As a result, the star reaches an equilibrium rotation law close to uniform
(see models in the above ref.), with always a strong thermal instability 
amplifying meridional circulation and thus chemical mixing.
Models show that  the high enrichments in magnetic stars with a dynamo 
(Sect. \ref{sectdyn}) are essentially due to the transport by 
meridional circulation. \emph{Thus, it is not the magnetic field itself which makes the mixing, but the thermal instability associated to the solid rotation created by the field.}

The high surface magnetic field of these stars, which likely have a significant mass loss,
produces a strong magnetic braking, which implies that  these stars will reach a rather low rotation velocities during their evolution. 
The braking would  
 tend to produce some internal differential rotation. However, the magnetic coupling 
 is certainly strong enough to maintain a rotation law close to uniform, as illustrated by the models.

\section{Dynamos in radiative layers: general properties and equations}  \label{sectdyn}

The major question concerning magnetic fields and stellar evolution is whether  a dynamo operates  in radiative zones of  differentially rotating stars. A magnetic field has  great consequences on the evolution of the rotation velocity  by exerting an efficient
 torque able  to impose a nearly uniform rotation. This influences all the model outputs (lifetimes, chemical abundances, tracks, chemical yields, supernova types)  as well as the rotation in the final stages, white dwarfs, neutron stars or black holes.
 
 Here we first examine some general equations implied by any  dynamo. The particular properties of the Tayler--Spruit (TS) dynamo  have been studied by \cite[Spruit (2002)]{Spruit02} and we are using many of the equations he derived. Spruit considered the 
 radiative zones in two cases, --1) when the $\mu$--gradient dominates, and --2)
 when the $\mu$--gradient is negligible. The more general equations of the TS dynamo have been developed  by (\cite[Maeder \& Meynet 2005]{Magn3}).
 The TS dynamo is at present a debated subject.
Some  numerical simulations by  \cite[Braithwaite (2006)]{Braithwaite06} and by   \cite[Brun et al. (2007)]{BrunZM07}   confirm the existence of Tayler's instability.
Braithwaite  also finds the existence of a dynamo loop in agreement with Spruit's analytical developments.
However, Zahn et al. do not find the dynamo loop proposed by Spruit and question
what may close the loop.

\subsection{Energy conservation}

If a dynamo is working in a differentially rotating radiative zone, it is governed by some general 
relations expressing the order of magnitude of its various  properties. 
First, the rate of magnetic energy production 
  $W_{\mathrm{B}}$ per unit of time and volume must be equal to the rate
  $W_{\nu}$ of  dissipation of rotational energy by the magnetic viscosity
  $\nu$. We assume here   that the whole energy 
  dissipated is converted into magnetic energy.
  The differential motions are those of  the shellular rotation with an angular velocity $\Omega(r)$, so that the velocity
  difference at radius $r$ is $dv= r \, d\Omega$.
  The amount of energy corresponding to a velocity difference $dv$ during a time $dt$  for an element of matter 
 $ dm$ in a volume $dV$ is 
  \begin{eqnarray}
W_\nu=\frac{1}{2} \, dm \,(dv)^2 \frac{1}{dV} \, \frac{1}{dt}= \frac{1}{2} {\varrho} \, \nu
\left(\frac{dv}{dr}\right)^2 = {1\over 2}\varrho\nu \Omega^2q^2 \quad \quad
\mathrm{with} \quad q = r \left|\nabla \Omega\right|/\Omega,
\label{Wnu}
\end{eqnarray}

\noindent
because the viscous time $dt$ over $dr$ is given by $dt=(dr)^2/\nu$.
The magnetic energy density is $u_{\mathrm{B}}={B^2}/(8 \pi)$, it
is produced within the characteristic growth time of the magnetic field $\sigma_{\mathrm{B}}^{-1}$, 
thus the rate $W_{\mathrm{B}}$ of magnetic energy creation by units of volume and time is
\begin{equation}
W_{\mathrm{B}}= \frac{B^2}{8 \pi} \sigma_{\mathrm{B}}= \frac{1}{2} \omega^2_{\mathrm{A}}r^2 \sigma_{\mathrm{B}}\varrho \;,
\label{WB}
\end{equation}

\noindent
where we have used   the expression of the Alfv\'en frequency 
$\omega_{\mathrm{A}} \, = \, \frac{B}{r (4 \, \pi \varrho)^{1/2}} $.
Now, let us  assume $W_{\nu}  =  W_{\mathrm{B}}$, i.e. that the excess of energy in the differential rotation
(compared to an average constant rotation) is converted to magnetic energy
by unit of time. This gives the following expression for the viscosity coefficient of magnetic coupling
\begin{eqnarray}
\nu = \frac{\omega^2_{\mathrm{A}}\, r^2 \, \sigma_{\mathrm{B}}}{\Omega^2 \,q^2} \; .
\label{nu1}
\end{eqnarray}

\noindent
This is the coefficient which intervenes in the expression for the transport
of angular momentum, in the Lagrangian form as given by Eq. (\ref{eqn7}) below.
Let us note that compared to the  energy available for the solar dynamo driven by convection, the amount of energy available from differential rotation is very limited.

\subsection{The $\alpha$ and $\omega$--effects: vertical instability and stretching of the field lines}

A dynamo needs both  the $\alpha$--effect and $\omega$--effect.
The $\alpha$--effect consists
in the generation of a poloidal field component from the horizontal component.
In the Sun, the $\alpha$--effect is created by the  convective motions and
by the twisting of the magnetic loop by the Coriolis force. However, other instabilities with a vertical component may produce the necessary $\alpha$--effect.
The $\omega$--effect consists mainly of the stretching of a small radial field component in the East--West directions. The winding--up of the field lines generates a stronger horizontal field component, converting some  kinetic energy  into magnetic energy.

If due to an instability in radiative layers, some  vertical displacements
(necessary for the $\alpha$--effect)
 with an  amplitude $l_r/2$ occur around an average stable position,
 the restoring buoyancy force produces  vertical oscillations with a
frequency equal to the Brunt--V\H{a}is\H{a}l\H{a} frequency $N$. 
The restoring oscillations will have an average density of kinetic energy 
$u_{\mathrm{N}}  =   f_{N} \, \rho \, \ell^2 \, N^2$,
where $f_{N} \sim 1$.
In order  to produce a vertical displacement,  the magnetic field must overcome the buoyancy effect. In terms of energy densities, this is
$ u_{\mathrm{B}} >  u_{\mathrm{N}} $, where $u_{\mathrm{B}}$ has been given in 
the previous section. 
 Otherwise the restoring force of gravity
 would  counteract the magnetic instability  at the dynamical timescale. From this
 condition, one obtains 
$\ell^2 <  \frac{1}{2f_{\mathrm{N}}} \,r^2 \, \frac{\omega^2_{\mathrm{A}}}{N^2}$.
 If, $f_{\mathrm{N}}= {1 \over 2}$, we  have the condition (\cite[Spruit 2002]{Spruit02})
 \begin{equation}
 \ell< l_r =  \; r \; \frac{\omega_{\mathrm{A}}}{N}  \; ,
 \label{lr}
 \end{equation}
 
  \noindent
  where  $r$ is the radius at the considered level in the star.

The stretching of the field lines for the $\omega$--effect is governed by the induction equation 

\begin{eqnarray}
\frac{\partial \vec{B}}{\partial t}= \vec{\nabla} \times (\vec{v}\times \vec{B}) +
\eta \, \nabla^2 \vec{B} \; .
\label{inducequ}
\end{eqnarray} 
\noindent
An  unstable vertical displacement of size $\ell$ from the azimuthal field of  lengthscale  $r$
and intensity $B_{\varphi}$ also feeds a radial field component $B_r$. The relative sizes of these two field components are
defined by the induction equation (\ref{inducequ}), which gives the following scaling over the time $\delta t$ characteristic 
of the unstable displacement,
\begin{eqnarray}
B_r \, \approx \,  \delta B  \, \approx \, \frac{1}{r} \,\frac{\ell}{\delta t} \, B_{\varphi} \,   \delta t  \; .
\end{eqnarray}

\noindent
For the maximum displacement $l_r$ given by Eq. \ref{lr}, this  gives
\begin{eqnarray}
\frac{B_r}{B_{\varphi}} \approx \frac{l_r}{r}  \; ,
\label{lrr}
\end{eqnarray}

\noindent
which provides (\cite[Spruit 2002]{Spruit02})  an estimate of the ratio of the radial to azimuthal fields.

\subsection{The magnetic and thermal  diffusivities}

 The magnetic diffusivity $\eta$ tends to damp the  instability, while
 the thermal diffusivity $K$ produces heat losses from the unstable fluid elements
and thus  reduces the buoyancy forces opposed to the magnetic instability.  Both effects
have to be accounted for.
 
 If the radial scale of the vertical instability is   small, the perturbation is quickly damped by the magnetic diffusivity  
  $\eta$ (in cm$^2$ s$^{-1}$). The radial amplitude  must satisfy,  
  \begin{eqnarray}
 l_r^2 >  \frac{\eta}{\sigma_{\mathrm{B}}} \; , 
 \label{lmin}
 \end{eqnarray} 
 
 \noindent
 where, as seen above, 
 $\sigma_{\mathrm{B}}$ is the characteristic frequency for the growth  of the  instability. 
The combination of the two limits   (\ref{lrr}) and (\ref{lmin}) gives for the case of marginal stability,
\begin{equation}
\eta \, = \, \frac{r^2 \, \omega^2_{\mathrm{A}} \, \sigma_{\mathrm{B}}}{N^2}   \; .
\label{premier}
\end{equation} 

\noindent
For given $\eta$ and $\sigma_{\mathrm{B}}$, this provides the minimum value of $\omega_{\mathrm{A}}$, and thus of the magnetic field $B$, for the instability to 
occur.
The instability is confined within a domain, limited on the large side 
by the stable stratification (\ref{lr}) and on the small scales by magnetic diffusion (\ref{lmin}). For the case of marginal stability, which is likely reached in evolution, this equation relates the magnetic diffusivity $\eta$ and the Alfv\'en frequency $\omega_{\mathrm{A}}$.\\

 The Brunt--V\"{a}is\"{a}l\"{a} frequency $N$  of a fluid element displaced
in a medium with account of both the magnetic and thermal diffusivities $\eta$ and
$K$ is  (\cite[Maeder \& Meynet 2004]{Magn2}),
\begin{eqnarray}
N^2 = \frac{\frac{\eta}{  K}} {\frac{\eta}{  K}  + \, 2} \, \; N^2_{T, \, \mathrm{ad}}+ N^2_{\mu}  \; , \label{N2} \\[2mm]
\mathrm{with} \quad  N ^2_{T,\,\mathrm{ad}}= \frac{g  \delta}{H_P} \left(
 \nabla_{\mathrm{ad}}-\nabla_{\mathrm{}} \right), \quad \mathrm{and}\quad N^2_{\mu}= \frac{g \, \varphi}{H_P} \nabla_{\mu} \,,
\label{Nfinal}
\end{eqnarray}
\noindent
The ratio $\eta/K$ of the magnetic to thermal diffusivities determines the
  heat losses. The factor of 2 is determined by the geometry of the instability, a factor of 2 applies to a thin slab, for a spherical element a factor of 6 is appropriate
 (\cite[Maeder \& Meynet 2005]{Magn3}). 

\subsection{The magnetic coupling and the timescale $\sigma_{\mathrm{B}}$}

  The  momentum of force $\vec{S}$ by volume unity due to the magnetic field  is obtained 
  by writing  the momentum of  the Lorentz force $\vec{F}_{\mathrm{L}}$.
  The current density $\vec{j}$ is given by the Maxwell equation $\frac{4 \, \pi}{c} \,\vec{j}= \vec{\nabla } \times {\vec{B }}$. Thus, one has 
  \begin{eqnarray}
  \vec{S}=\vec{r} \times \vec{F}_{\mathrm{L}}=\frac{1}{c} \vec{r} \times(\vec{j} \times \vec{B})
  =\frac{1}{4  \, \pi} \vec{r} \times \left((\vec{\nabla} \times \vec{B}) \times \vec{B} \right) \, , \; \;\\[2mm]
  \mathrm{in \; modulus} \quad S \, \approx \,  \frac{1}{4 \; \pi} \; B_{\mathrm{r}} B_{\varphi} \; = \;
  \frac{1}{4 \; \pi} \;  \left(\frac{l_{\mathrm{r}}}
  {r}\right) B_{\varphi}^2 = 
  \; \rho \; r^2 \; \left(\frac{\omega_{\mathrm{A}}^3}{N}\right)  \; .
  \label{S}
  \end{eqnarray}
 
 \noindent
 The units of $S$ are g s$^{-2}$ cm$^{-1}$, the same as for $B^2$ in the Gauss system.
 The kinematic viscosity $\nu$ (in cm$^2$ s$^{-1}$) 
 for the vertical transport of angular momentum is
 \begin{eqnarray}
 \nu=\frac{\eta}{\varrho} = \frac{1}{\varrho}  \, F  \, \frac{dr}{dv}=
 \frac{1}{\varrho} \, F \, \frac{dr}{r d\Omega} = 
 \frac{1}{\varrho} \, F \, \frac{d \ln r}{\Omega \,d \ln \Omega} \;,
 \end{eqnarray}
  
 \noindent
where  F   is a force by surface unity, which also
 corresponds to a momentum of force by volume unity in
   g s$^{-2}$ cm$^{-1}$. $\vec{F}$ is  applied horizontally to a slab of velocity $v$ in
   a direction perpendicular to $r$. 
 Considering only positive quantities, with  $q= \left|d\ln \Omega/d \ln r\right|$,
 one has
 \begin{equation}
 \nu = \frac{S}{\rho \; q \; \Omega} = \; \frac{\omega^3_{\mathrm{A}} \, r^2}{N\, q \, \Omega} \;.
 \label{nu2}
 \end{equation}
 
 Now, we can compare this expression for $\nu$ to Eq. (\ref{nu1}) and  get
 \begin{eqnarray}
 \sigma_{\mathrm{B}} \; = \; \frac{\omega_{\mathrm{A}} \, \Omega \, q}{N} \;.
 \label{sigfdt} 
 \end{eqnarray}
 \noindent
 This important expression  relates the growth rate of the magnetic field to 
 its amplitude (through $\omega_{\mathrm{A}}$). It can also be obtained  by expressing the amplification time 
 $\tau_{\mathrm{a}}$ of the field line $B_r$ to the level of $B_{\varphi}$ by the winding--up of the field line $B_{\varphi} \, \approx \, B_r \, r \,\left( - \frac{\partial \Omega}{\partial r}\right) \, \tau_a $,
 \begin {eqnarray}
\tau_{\mathrm{a}} \, = \,\frac{ N} {\omega_{\mathrm{A}} \Omega \, q} \; ,
\end{eqnarray}

\noindent
and equaling this to $\sigma^{-1}_{\mathrm{B}}$ we also get Eq. (\ref{sigfdt}).

\subsection{The basic equations}
Introducing the expression (\ref{sigfdt}) of $\sigma_{\mathrm{B}}$ in Eq. (\ref{premier}), we get
 \begin{eqnarray}
 \eta \,=\, r^2 \, \Omega \, q \left(\frac{\omega_{\mathrm{A}}}{N}\right)^3 \, \, .
 \label {1etaN}
 \end{eqnarray}
 \noindent
 Also, with the expression (\ref{N2}), we can write for $\sigma^2_{\mathrm{B}}$
 \begin{eqnarray}
 \sigma^2_{\mathrm{B}} \, = \,\frac{\omega^2_{\mathrm{A}} \, \Omega^2 \, q^2}{\frac{\frac{\eta}{  K}} {\frac{\eta}{  K}  + \, 2} \, \; N^2_{T, \, \mathrm{ad}}+ N^2_{\mu}  } \;.
 \label{2sigmaN}
 \end{eqnarray} 
\noindent
These equations are quite general. If the growth rate $\sigma_{\mathrm{}}$ of the 
instability is known, the two equations (\ref{1etaN}) and (\ref{2sigmaN}) form a system of 2 equations with 2 unknowns $\eta$  and $\omega_{\mathrm{A}}$.

\section{The case of Tayler--Spruit dynamo}

In a non--rotating star the growth rate of the Tayler instabilty is the Alfv\'en frequency $\omega_{\mathrm{A}}$. In a rotating star, the instability is also present,  however the characteristic growth rate $\sigma_{\mathrm{B}}$ of the instability is, if $\omega_{\mathrm{A}} \ll \Omega$,
\begin{eqnarray}
\sigma_{\mathrm{B}} \, = \, \frac{\omega_{\mathrm{A}}^2}{\Omega}  \; ,
\label{sigcoriolis}
\end{eqnarray}

\noindent
 because the growth rate of the instability  is reduced by the Coriolis force (\cite[Spruit 2002]{Spruit02}). If so, Eqs. (\ref{1etaN}) and  (\ref{2sigmaN}) become
 \begin{eqnarray}
\left(\frac{\omega_{\mathrm{A}}}{\Omega}\right)^2  =
\frac{\Omega^2 \; q^2}
{  N^2_{T, \, \mathrm{ad}} \; \frac{\eta / K} {\eta / K  \;+ \; 2} + N^2_{\mu}},
\label{deusse} \\[2mm]
\eta \;= \; \frac{r^2 \; \Omega}{q^2} \; \left( \frac{\omega_{\mathrm{A}}}
{\Omega}\right)^6  \; .
\label{etaa}
\end{eqnarray}

\noindent
This  forms a  system of 2 equations for the 2  unknown 
quantities $\eta$ and $\omega_{\mathrm{A}}$.
 With a new variable 
 $x=\left({\omega_{\mathrm{A}}}/{\Omega}\right)^2$, we  get  (\cite[Maeder \& Meynet 2005]{Magn3}) a system of degree 4,
\begin{eqnarray}
\frac{r^2 \Omega}{q^2 K} \left(N_{\mathrm{T}}^2 + N_{\mu}^2 \right)  x^4-
\frac{r^2 \Omega^3}{K} x^3 + 2 N_{\mu}^2 \; x - 2 \Omega^2 q^2 = 0 \; .
\label{equx}
\end{eqnarray}

\noindent 
 The solution $x$
  provides  the value of the Alfv\'en frequency $\omega_{\mathrm{A}}$ and thus of the $\vec{B}$ field. By
 (\ref{etaa}) one gets the value of  $\eta$ and by (\ref{nu2}) the value of $\nu$. The above equation applies to the general 
case where both $N_{\mu}$ and $N_{\mathrm{T}}$ are different from zero
and where thermal losses may reduce the restoring buoyancy force. The solutions of this equation have been discussed (\cite[Maeder \& Meynet 2005]{Magn3}).
In particular,  if $N_T=0$, one has
\begin{eqnarray}
x = \left(q \; \frac{\Omega}{N_{\mu}}\right)^2   \quad \mathrm{and} \quad
\eta = r^2 \Omega  q^4 \; \left(\frac{\Omega}{N_{\mu}}\right)^6  \; ,
\label{etaz0}
\end{eqnarray}
\noindent
which shows that the mixing of chemical elements decreases
 strongly for  larger $\mu$ gradients and grows fast for larger $q$ values.

\begin{figure}[!t]
\centering
\includegraphics[height=6.0cm,width=6.2cm]{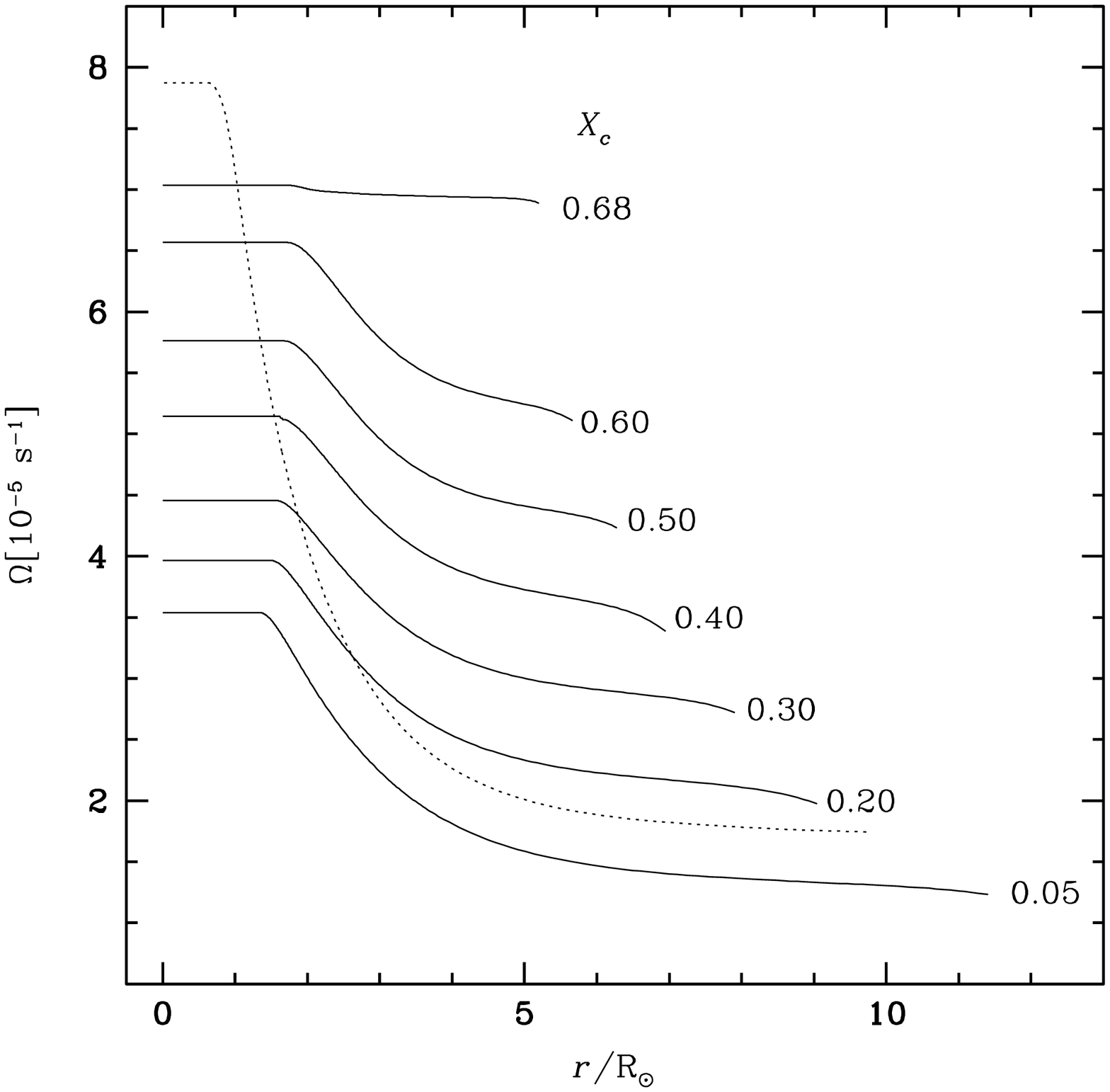}
\includegraphics[height=6.0cm,width=6.2cm]{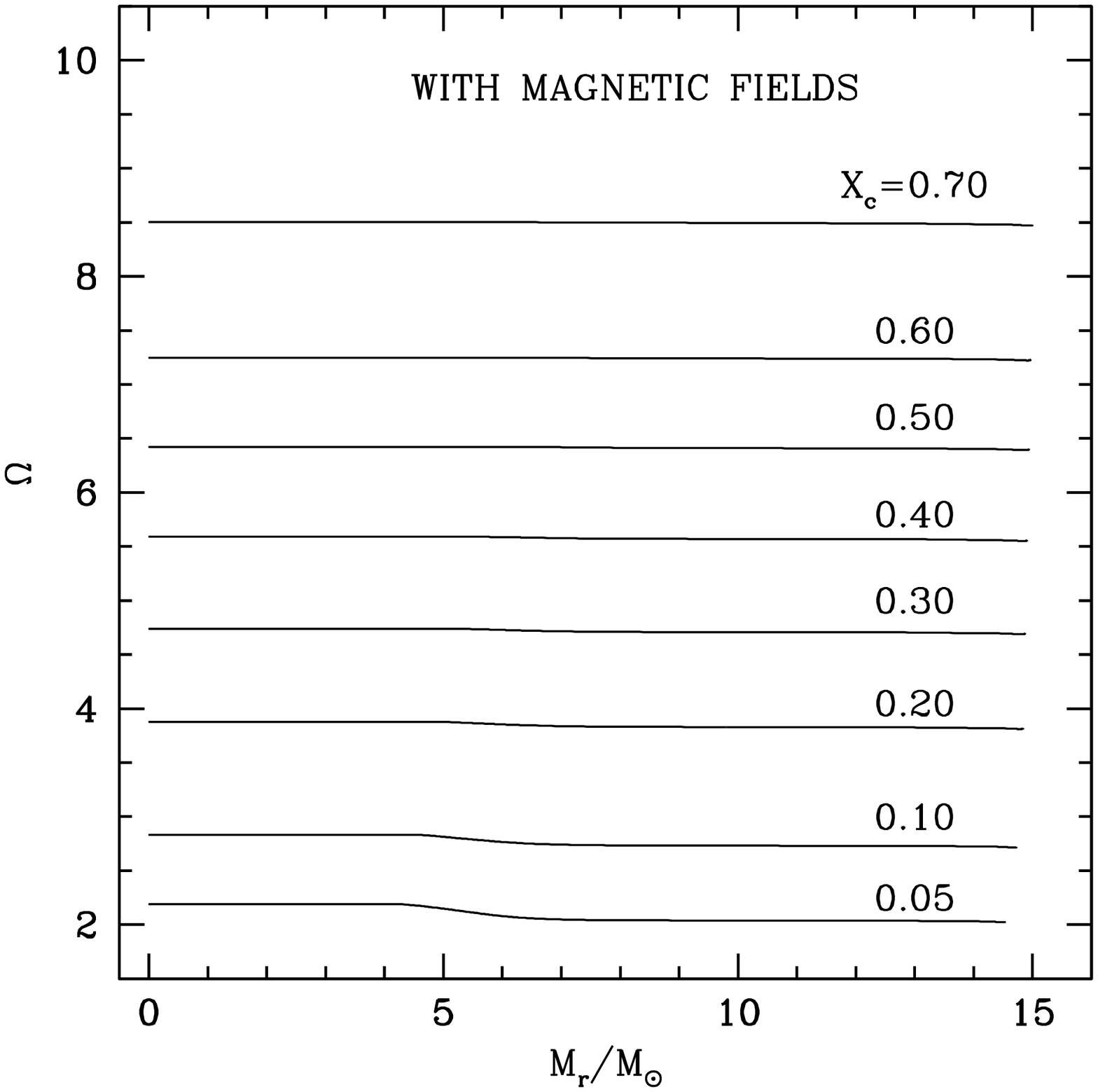}
\caption{Left: evolution of the angular velocity $\Omega$ 
 as a function of the distance to the center
in a 20 M$_\odot$ star with $v_{\rm ini}$ = 300 km s$^{-1}$. 
$X_c$ is the hydrogen mass fraction at the center.
The dotted line shows the profile when the He--core contracts at the end
of the H--burning phase (\cite[Meynet \& Maeder 2000]{MMV}). Right: rotation profiles
at various stages of evolution  (labeled by the central H content $X_{\mathrm{c}}$) of a 15 M$_{\odot}$ 
model with $X=0.705, Z=0.02$, an initial velocity of 300 km s$^{-1}$ and magnetic field from the TS dynamo
(\cite[Maeder \& Meynet 2005]{Magn3}).}
\label{Fig.2}
\end{figure}  
 The ratio $\omega_{\mathrm{A}}/\Omega$ given by the solution of (\ref{equx}) has to be equal or larger than 
 the minimum value defined by (\ref{premier}). This  leads to a condition on the minimum differential rotation  for the
 dynamo to work (\cite[Spruit 2002]{Spruit02}),
 \begin{eqnarray}
q \;  > \; \left({N \over \Omega}\right)^{7/4} \left({\eta \over r^2 N}\right)^{1/4}\; ,
\label{cond1}
\end{eqnarray}

\noindent
 When $N^2$ is
larger, as for example when there is a significant $\mu$ gradient, the differential rotation necessary for the dynamo to operate must also be larger.
 If the above condition is not fulfilled, there is  no stationary solution and the dynamo does not operate.
In practice,  this  often occurs in the outer stellar envelope.

\subsection{Equations of transport of chemical elements and angular momentum}
The equation for the transport of chemical species 
with mass fractions $X_i$ is at a Lagrangian mass coordinate $M_r$,
\begin{eqnarray}
	\varrho{\partial X_i \over \partial t}= {1 \over r^2} \, {\partial \over \partial r}
\left (\varrho \, r^2 \, (D_{\mathrm{eff}}+\eta) \, {\partial X_i \over \partial r}\right )  \; ,
\label{difx}
\end{eqnarray}
\noindent
where $D_{\mathrm{eff}}$ is the  coefficient for the transport
by meridional circulation and the possible horizontal turbulence.
The equation for the transport of   angular momentum is
\begin{eqnarray}
\varrho \, {\partial \over \partial t}(r^2 \overline \Omega)_{M_r}={1 \over 5 \, r^2}{\partial \over \partial r}(\varrho \,  r^4 \overline \Omega \,U_2(r))
+{1 \over r^2}{\partial \over \partial r}\left(\varrho  \, \nu \,  r^4 \,  {\partial \overline \Omega \over \partial r} \right) \; ,
\label{eqn7}
\end{eqnarray}

\noindent
where $U_2(r)$ is the amplitude of the radial component of the velocity of meridional circulation and $\nu$ the value given by (\ref{nu1}).
 This equation   is currently applied in stellar models for calculating the evolution of $\Omega$. With account of the detailed expression of $U_2(r)$, which contains 
terms up to the third spatial derivative of $\Omega(r,t)$, the above equation is 
of the fourth order  and  its numerical solution requires great care.

\begin{figure}[!t]
\centering
\includegraphics[angle=-90,width=12cm]{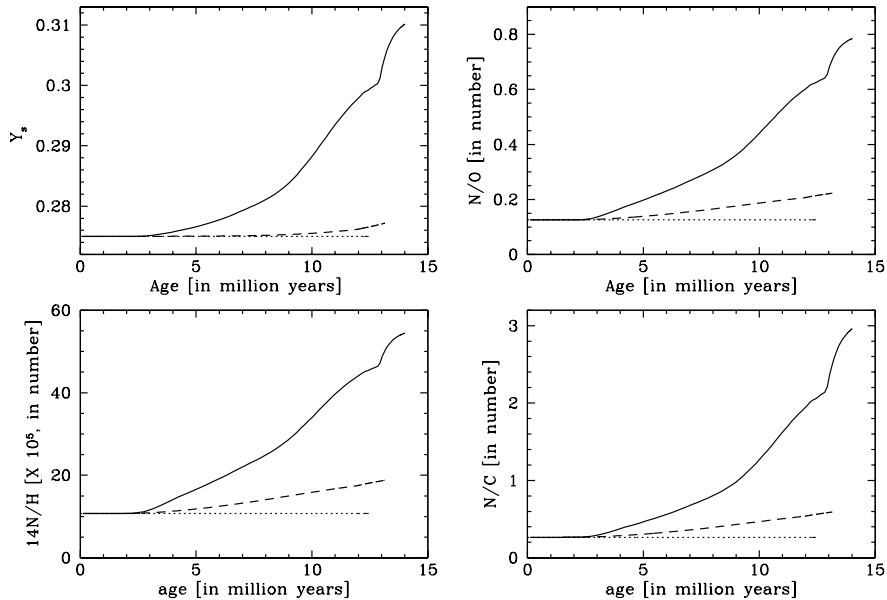}
\caption{Time evolution of the surface helium content $Y_{\mathrm{s}}$ in mass fraction, of the 
  N/O, N/H and N/C in mass fraction for various models: The dotted line
  applies to the  model without rotation, the short--broken line to the model with rotation 
  ($\upsilon_{\rm ini}$= 300 km s$^{-1}$) but without
  magnetic fields, the continuous 
  line to the model with rotation ($\upsilon_{\rm ini}$= 300 km s$^{-1}$) and magnetic fields from the TS dynamo (\cite[Maeder \& Meynet 2005]{Magn}).}
\label{Fig4}
\end{figure}

 \section{Numerical models}  \label{num}
 
 Numerical models accounting for  meridional circulation and magnetic field generated  by the TS dynamo have been computed (\cite[Maeder \& Meynet 2005]{Magn}).  The resulting  fields are  a few $10^4$ G  
   through most of the  envelope, with the exception of the outer layers
   where differential rotation is too small to sustain the TS dynamo.
   The diffusion coefficient for the transport of angular momentum is 
    large. In the Sun, it is of the order of $10^2$ to $10^6$ cm$^2$ s$^{-1}$, sufficient to impose solid body rotation at the age of the Sun (\cite[Eggenberger et al. 2005]{Eggenb05}). This coefficient
   is much larger in more massive stars, 
    in the range of $10^{10}$ to $10^{12}$ cm$^2$ s$^{-1}$ in a 15 M$_{\odot}$ star. There, it imposes nearly solid body rotation during most of the MS phase,
   while without the field there is a high differential rotation (Fig. \ref{Fig.2}).

    The nearly solid body rotation of star with magnetic field
    drives meridional circulation currents which are  faster than the currents   in differentially rotating stars.
    This  leads 
    to large surface enrichments in N and He together with C,O depletions in massive stars (Fig. \ref{Fig4}). Thus, the enhanced mixing results  from the thermal  instability enhanced by uniform rotation.
     The stellar lifetimes are enlarged by the mixing and the other model
     outputs are also modified  (\cite[Maeder \& Meynet 2005]{Magn}).
 Therefore, magnetic field is also a basic
ingredient of stellar evolution.

\end{document}